\documentstyle[12pt]{article}
\begin{document}
\title{THE ANOMALOUS BEHAVIOUR OF FERMIONS IN A THIN WIRE}
\author{B.G. Sidharth$^*$\\
Centre for Applicable Mathematics \& Computer Sciences\\
B.M. Birla Science Centre, Adarsh Nagar, Hyderabad - 500063 (India)}
\date{}
\maketitle
\footnotetext{$^*$E-mail:birlasc@hd1.vsnl.net.in}
\begin{abstract}
Motivated by the recent development of insulated nano-tubes and the attempts
to develop conducting nano wires in such tubes, we examine the Fermionic
behaviour in extremely thin wires. Although the one-dimensional problem
has been studied in detail over the years, it is an extreme idealization:
We consider the more realistic scenario of thin wires which are nevertheless
three dimensional. We show that the assembly of Fermions behaves as if it
is below the Fermi temperature, and in the limit of one dimension, in the
ground state as well. Thus there are indeed Bosonization features. These
conclusions are checked from an independent stand point.
\end{abstract}
\section{Introduction}
The problem of Fermions in one dimension has been studied in detail over
the years by Tomonaga, Takahashi and several others\cite{r1,r2,r3,r4} ever
since Bloch suggested that in an approximate sense a Fermi gas can be
described in terms of a quantized field of sound waves. However it must
be remembered that the one dimensional case is an extreme generalization
because Fermions as we know are meaningful only in three dimensions\cite{r5,r6}.
That is why, the well known one dimensional Lorentz covariant equation (in
the Weyl notation) viz.,
$$(\imath^\mu \gamma^\mu \partial_\mu - \frac{mc}{\hbar}) \psi = 0 (\mu = 1,2)$$
exhibits neutrino like handedness, but not spin (cf.ref.\cite{r7} for a
rationale). However it is reasonable to suppose that the detailed one
dimensional analysis would still be an approximate description of a thin
wire.\\
We now consider the problem in a more realistic context, that of an
extremely thin, though strictly speaking three dimensional wire. One
motivation for this is the fact that very recently a team from Connissariat
l' Energie Atomique of France has been able to develop nano tubes of
insulating material and attempts are on to develop conducting narrow wires
in these tubes\cite{r8}.\\
In Section 2 we shall show the interesting result that the Fermions will
exhibit either a Bosonization or they will behave like an assembly whose
energy is below the Fermi energy. In Section 3 we justify this conclusion
from the point of view of the one dimensional theory and further show that
the assembly also behaves as if it is in the ground state. In Section 4 we
will show that an assembly below the Fermi energy leads back to Bosonization
features. Finally in Section 4 we will make a few brief observations.
\section{Fermions in a Thin Wire}
Strictly speaking in this case we loose the exact solvability available in
the exact one dimensional case. However we could use the fact that Tomonaga's
original conclusion that the one dimensional assembly is described by sound
quanta together with weakly interacting Fermions has since been extended
to three dimensions (cf.ref.\cite{r3}). We could also use the Sommerfeld
non interacting Fermions in a box or Landau's Fermi liquid model in which
there is one to one correspondence to non interacting Fermions (cf.ref.\cite{r4}).\\
We now consider a free electron gas in an extremely thin metallic wire\cite{r9}.
We first take up the case when the dispersion in momentum is not large. That
is, we are considering the one dimensional case wherein the entire collection
of Fermions has more or less the same momentum and therefore energy.\\
The occupation number of Fermions is then given by (cf.ref.\cite{r10}):
\begin{equation}
n_{\vec p} = \frac{1}{z^{-1}\epsilon^{\beta \epsilon_p}+1}\label{e1}
\end{equation}
$$(\beta \equiv 1/kT)$$
The total number of Fermions in a volume $V$ is given by,
\begin{equation}
N = \sum_{\vec p} n_{\vec p}\label{e2}
\end{equation}
As usual, we replace $\sum_{\vec p}$ by $\frac{V}{h^3} \int d^3p.$ Further,
for an extremely thin "one dimensional volume" e.g., wire, (in space and
therefore also in momentum space) if the cross-section is $\sigma (<<1),$
then $d^3p = \sigma dp.$ Using these facts and also (\ref{e1}) in (\ref{e2})
we get,
$$N = \frac{V\sigma}{h^3}\int \frac{dp}{z^{-1}\epsilon^{\beta \epsilon_p}+1}
< \frac{V \sigma}{h^3} \int \frac{zdp}{e^{\beta \epsilon_p}}$$
or
\begin{equation}
1 < \frac{\nu \sigma}{h^3} \int z \frac{dp}{e^{\beta \epsilon_p}} \sim
\frac{\nu \sigma z}{\lambda^3}\label{e3}
\end{equation}
where $\nu = \frac{V}{N}$ and $\lambda = (\frac{2\pi \hbar^2}{mkT})^{1/2}
\sim \frac{\hbar}{(mkT)^{1/2}}$, which is of the order of the de Broglie
wavelength.\\
We first consider the case where neither is $z$ not very much larger than
$1$ nor $\frac{\lambda^3}{\nu}$ is very much smaller than $1$.\\
So, the term on the right side of inequality (\ref{e3}) is of the order of
$\sigma$, which is vanishingly small. This is a contradiction. The
contradiction disappears if, instead of (\ref{e1}), we use the Bosonic
occupation number,
$$n_{\vec p} = \frac{1}{z^{-1}\epsilon^{\beta \epsilon_p}-1}$$
In that case the inequality (\ref{e3}) is reversed and there is no problem.
That is, the one dimensional Fermions apparently obey Bose-Einstein statistics.
If $\frac{\lambda^3}{\nu} < < 1 \mbox{then} \quad z \sim \frac{\lambda^3}{\nu}$
\cite{r10} and the contradiction persists.\\
If on the other hand $z > > 1 \mbox{then} \quad \lambda^3/\nu \sim (log z)^{3/2}$
which corresponds to the degenerate case where the temperature is below
the Fermi temperature.\\
The above cases were considered under the assumption that the dispersion
of momentum is not very large. If this dispersion is very large, then
$\Delta p > > 1,$ also and $\lambda$ is very small. This corresponds to a
high velocity or momentum of the Fermions: Specifically this leads to the
case where we have a low density and a high temperature distribution, in
which case Fermi-Dirac and Bose-Einstein statistics both converge to the
Boltzmann distribution.\\
So we have either an anomalous Bosonic type behaviour or the temperature
is below the Fermi temperature. We examine this more closely now.
\section{One Dimensional Theory}
In the strictly one dimensional case we have the Fermion-Bosonic Transmutations
(cf.ref.\cite{r4}), with the Fermionic partition function coinciding with
the Bosonic one. Further, the average energy per unit length is given by
\begin{equation}
e = \frac{\pi (kT)^2}{6\hbar \nu_F}\label{e4}
\end{equation}
where $\nu_F \equiv \hbar \pi (N/L)/m, L$ being the length of the one
dimensional wire and $N$ the number of Fermions therein. This is the one
dimensional version of the Stephan Boltzmann law for radiation\cite{r11}. Denoting the
average interparticle distance,
$$\frac{L}{N} \equiv (\nu)^{1/3},$$
and using the fact that\cite{r10}
$$kT_F = (\frac{\hbar^2}{2m}) (\frac{6\pi^2}{\nu})^{2/3},$$
and remembering that,
$$kT = e\nu^{1/3},$$
we can easily deduce from (\ref{e4}) that,
$$T = \frac{3}{5} T_F$$
Interestingly this not only shows that the temperature is below the Fermi temperature,
but also that the gas is in the ground state\cite{r10}, whatever be the
temperature.
\section{Below the Fermi Temperature}
Finally we show that the collection of Fermions below the Fermi temperature,
exhibits a Bosonization in character. As is known the Fermi energy is given
by
\begin{equation}
\epsilon_F = p^2_F/2m = (\frac{\hbar^2}{2m}) (\frac{6\pi^2}{\nu})^{2/3}\label{e5}
\end{equation}
where $\nu^{1/3}$ is the interparticle distance. On the other hand, in a
different context, for phonons, the maximum frequency is given by,
(cf.ref.\cite{r10}),
\begin{equation}
\omega_m = c(\frac{6\pi^2}{\nu})^{1/3}\label{e6}
\end{equation}
This occurs for the phononic wavelength $\lambda_m \approx$ inter-atomic
distance between the atoms, $\nu^{1/3}$ being, again, the mean distance between
the phonons. '$c$' in (\ref{e6}) is the velocity of the wave, the velocity
of sound in this case. The wavelength $\lambda_m$ is given by,
$$
\lambda_m = \frac{2\pi c}{\omega_m}
$$
We can now define the momentum $p_m$ via the de Broglie relation,
$$\lambda_m = \frac{h}{p_m},$$
which gives,
\begin{equation}
p_m = \frac{\hbar}{c}\omega_m, \hbar \equiv \frac{h}{2\pi}\label{e7}
\end{equation}
We can next get the maximum energy corresponding to the maximum frequency
$\omega_m$ given by (\ref{e6}),
\begin{equation}
\epsilon_m = \frac{p^2_m}{2m} = \frac{\hbar^2}{2m}(\frac{6\pi^2}{\nu})^{2/3}\label{e8}
\end{equation}
\begin{equation}
n_{\vec p} = \frac{1}{z^{-1}\epsilon^{\beta \epsilon_p}+1}\label{e1}
\end{equation}
$$(\beta \equiv 1/kT)$$
The total number of Fermions in a volume $V$ is given by,
\begin{equation}
N = \sum_{\vec p} n_{\vec p}\label{e2}
\end{equation}
As usual, we replace $\sum_{\vec p}$ by $\frac{V}{h^3} \int d^3p.$ Further,
for an extremely thin "one dimensional volume" e.g., wire, (in space and
therefore also in momentum space) if the cross-section is $\sigma (<<1),$
then $d^3p = \sigma dp.$ Using these facts and also (\ref{e1}) in (\ref{e2})
we get,
$$N = \frac{V\sigma}{h^3}\int \frac{dp}{z^{-1}\epsilon^{\beta \epsilon_p}+1}
< \frac{V \sigma}{h^3} \int \frac{zdp}{e^{\beta \epsilon_p}}$$
or
\begin{equation}
1 < \frac{\nu \sigma}{h^3} \int z \frac{dp}{e^{\beta \epsilon_p}} \sim
\frac{\nu \sigma z}{\lambda^3}\label{e3}
\end{equation}
where $\nu = \frac{V}{N}$ and $\lambda = (\frac{2\pi \hbar^2}{mkT})^{1/2}
\sim \frac{\hbar}{(mkT)^{1/2}}$, which is of the order of the de Broglie
wavelength.\\
We first consider the case where neither is $z$ not very much larger than
$1$ nor $\frac{\lambda^3}{\nu}$ is very much smaller than $1$.\\
So, the term on the right side of inequality (\ref{e3}) is of the order of
$\sigma$, which is vanishingly small. This is a contradiction. The
contradiction disappears if, instead of (\ref{e1}), we use the Bosonic
occupation number,
$$n_{\vec p} = \frac{1}{z^{-1}\epsilon^{\beta \epsilon_p}-1}$$
In that case the inequality (\ref{e3}) is reversed and there is no problem.
That is, the one dimensional Fermions apparently obey Bose-Einstein statistics.
If $\frac{\lambda^3}{\nu} < < 1 \mbox{then} \quad z \sim \frac{\lambda^3}{\nu}$
\cite{r10} and the contradiction persists.\\
If on the other hand $z > > 1 \mbox{then} \quad \lambda^3/\nu \sim (log z)^{3/2}$
which corresponds to the degenerate case where the temperature is below
the Fermi temperature.\\
The above cases were considered under the assumption that the dispersion
of momentum is not very large. If this dispersion is very large, then
$\Delta p > > 1,$ also and $\lambda$ is very small. This corresponds to a
high velocity or momentum of the Fermions: Specifically this leads to the
case where we have a low density and a high temperature distribution, in
which case Fermi-Dirac and Bose-Einstein statistics both converge to the
Boltzmann distribution.\\
So we have either an anomalous Bosonic type behaviour or the temperature
is below the Fermi temperature. We examine this more closely now.
\section{One Dimensional Theory}
In the strictly one dimensional case we have the Fermion-Bosonic Transmutations
(cf.ref.\cite{r4}), with the Fermionic partition function coinciding with
the Bosonic one. Further, the average energy per unit length is given by
\begin{equation}
e = \frac{\pi (kT)^2}{6\hbar \nu_F}\label{e4}
\end{equation}
where $\nu_F \equiv \hbar \pi (N/L)/m, L$ being the length of the one
dimensional wire and $N$ the number of Fermions therein. This is the one
dimensional version of the Stephan Boltzmann law for radiation\cite{r11}. Denoting the
average interparticle distance,
$$\frac{L}{N} \equiv (\nu)^{1/3},$$
and using the fact that\cite{r10}
$$kT_F = (\frac{\hbar^2}{2m}) (\frac{6\pi^2}{\nu})^{2/3},$$
and remembering that,
$$kT = e\nu^{1/3},$$
we can easily deduce from (\ref{e4}) that,
$$T = \frac{3}{5} T_F$$
Interestingly this not only shows that the temperature is below the Fermi temperature,
but also that the gas is in the ground state\cite{r10}, whatever be the
temperature.
\section{Below the Fermi Temperature}
Finally we show that the collection of Fermions below the Fermi temperature,
exhibits a Bosonization in character. As is known the Fermi energy is given
by
\begin{equation}
\epsilon_F = p^2_F/2m = (\frac{\hbar^2}{2m}) (\frac{6\pi^2}{\nu})^{2/3}\label{e5}
\end{equation}
where $\nu^{1/3}$ is the interparticle distance. On the other hand, in a
different context, for phonons, the maximum frequency is given by,
(cf.ref.\cite{r10}),
\begin{equation}
\omega_m = c(\frac{6\pi^2}{\nu})^{1/3}\label{e6}
\end{equation}
This occurs for the phononic wavelength $\lambda_m \approx$ inter-atomic
distance between the atoms, $\nu^{1/3}$ being, again, the mean distance between
the phonons. '$c$' in (\ref{e6}) is the velocity of the wave, the velocity
of sound in this case. The wavelength $\lambda_m$ is given by,
$$
\lambda_m = \frac{2\pi c}{\omega_m}
$$
We can now define the momentum $p_m$ via the de Broglie relation,
$$\lambda_m = \frac{h}{p_m},$$
which gives,
\begin{equation}
p_m = \frac{\hbar}{c}\omega_m, \hbar \equiv \frac{h}{2\pi}\label{e7}
\end{equation}
We can next get the maximum energy corresponding to the maximum frequency
$\omega_m$ given by (\ref{e6}),
\begin{equation}
\epsilon_m = \frac{p^2_m}{2m} = \frac{\hbar^2}{2m}(\frac{6\pi^2}{\nu})^{2/3}\label{e8}
\end{equation}
Comparing (\ref{e5}) and (\ref{e8}), we can see that $\epsilon_m$ and
$p_m$ exactly correspond to $\epsilon_F$ and $p_F$.\\
The Fermi energy in (\ref{e5}) is obtained as is known by counting all energy
levels below the Fermi energy $\epsilon_F$ using Fermi-Dirac statistics,
while the maximum energy in (\ref{e8}) is obtained by counting all energy
levels below the maximum value, but by using Bose-Einstein statistics
(cf.ref.\cite{r10}).\\
We can see why inspite of this, the same result is obtained in both cases.
In the case of the Fermi energy, all the lowest energy levels below $\epsilon_F$
are occupied with the Fermionic occupation number $\langle n_p \rangle = 1,
p <p_F.$ Then, the number of levels in a small volume about $p$ is $d^3p$.
This is exactly so for the Bosonic levels also. With the correspondence given
in (\ref{e7}), the number of states in both cases coincide and it is not
surprising that (\ref{e5}) and (\ref{e8}) are the same.\\
In effect, Fermions below the Fermi energy have a strong resemblance to
bosonic phonons.\\
All this is in agreement with the considerations of Section 2 and Section 3.
\section{Discussion}
The Bosonization encountered above should reflect in an anomalous behaviour.
Indeed the experimentally observed superfluidity of $He^3$ does exhibit
inexplicable features, even though the phenomenon is sought to be explained
in terms of Cooper pairs of the BCS Theory\cite{r12}. Indeed in a preliminary
communication\cite{r13}, it had been suggested on the basis of similar
arguments that electrons in an extremely thin wire could display a temperature
independent superconductivity type phenomenon.\\
It is interesting to note that from a completely different point of view, the
Bosonization of Fermions in low dimensions or at low energies has been suggested (cf.ref.\cite{r7} and
ref.\cite{r14,r15}). This is because low energies imply length scales
much greater than the Compton wavelength and according to the model discussed
in\cite{r14}, at such scales the behaviour would be Bosonic. Indeed this is
also suggested by the well known fact that in Quantum Field Theory the spin
statistics connection, according to which Bosons are quantized by commutators
as distinct from Fermions which are quantized by anti-commutators is strictly
speaking true only at wavelengths not much greater than the Compton
wavelength\cite{r16}.
\section{Addendum}
Very recent experimental results on carbon nanotubes[17,18,19,20] exhibit
the one dimensional nature of conduction and behaviour like low temperature
quantum wires thus confirming the results discussed.

\end{document}